\documentstyle[preprint,aps]{revtex}
\tighten
\draft
\widetext
\input epsf
\preprint{HUTP-98/A043, NUB 3179}
\bigskip
\bigskip
\begin{document}
\title{Anomaly Free Non-Supersymmetric Large $N$ Gauge Theories\\
from Orientifolds}
\medskip
\author{Zurab Kakushadze\footnote{E-mail: 
zurab@string.harvard.edu}}
\bigskip
\address{Lyman Laboratory of Physics, Harvard University, Cambridge, 
MA 02138\\
and\\
Department of Physics, Northeastern University, Boston, MA 02115}
\date{June 11, 1998}
\bigskip
\medskip
\maketitle

\begin{abstract}
{}We construct anomaly free non-supersymmetric large $N$ gauge theories from orientifolds
of Type IIB on ${\bf C}^3/\Gamma$ orbifolds. In particular, massless as well as tachyonic 
one-loop tadpoles are cancelled in these models. This is achieved by starting with ${\cal N}=1,2$ supersymmetric orientifolds with well defined world-sheet description and including discrete torsion (which breaks supersymmetry) in the orbifold action. In this way we obtain non-trivial non-chiral as well as anomaly free chiral large $N$ gauge theories. We point out certain subtleties
arising in the chiral cases. Subject to certain assumptions, these theories are shown to have the property that computation of any $M$-point correlation function in these theories reduces to
the corresponding computation in the parent ${\cal N}=4$ oriented theory. This generalizes the analogous results recently obtained in supersymmetric large $N$ gauge theories from orientifolds, as well as in (non)supersymmetric large $N$ gauge theories without orientifold planes.  
\end{abstract}

\section{Introduction}

{}Recent developments in the AdS/CFT correspondence (see, {\em e.g.}, \cite{Kleb,Gubs,Mald,Poly,Witt}) motivated
a set of conjectures proposed in \cite{KaSi,LNV} which state that certain gauge theories with
${\cal N}=0,1$ supersymmetries are (super)conformal. These gauge theories are constructed by 
starting from a $U(N)$ gauge theory with ${\cal N}=4$ space-time supersymmetry in four dimensions, and orbifolding by a finite discrete subgroup $\Gamma$ of the $R$-symmetry group
$Spin(6)$ \cite{LNV}. These conjectures were shown at one-loop level for ${\cal N}=0$ theories
\cite{KaSi,LNV}, and to two loops for ${\cal N}=1$ theories using ordinary field theory techniques \cite{LNV}. 

{}In the subsequent development \cite{BKV} these conjectures were shown to be correct to
all loop orders in the large $N$ limit of 't Hooft \cite{thooft} (also see \cite{BJ}\footnote{For other related works, see, {\em e.g.}, \cite{Iban,HSU}.}). 
In particular, in \cite{BKV} the above gauge theories were obtained in the $\alpha^\prime\rightarrow 0$ limit of Type IIB with $N$ parallel D3-branes imbedded in an orbifolded space-time. The work in \cite{BKV} was generalized in \cite{zura} where 
the Type IIB string theory included both D3-branes and orientifold 3-planes (with the transverse space being ${\bf C}^3/\Gamma$). In certain cases string consistency also requires presence of D7-branes and orientifold 7-planes. This corresponds to Type IIB orientifolds. Introducing orientifold planes
is necessary to obtain $SO$ and $Sp$ gauge groups (without orientifold planes the gauge group
is always unitary), and also allows for additional variety in possible matter content. 

{}In the presence of orientifold planes the string world-sheet topology is characterized by 
the numbers $b$ of boundaries (corresponding to D-branes), 
$c$ of cross-caps (corresponding to orientifold planes), and $g$ of handles (corresponding to closed string loops). Such a 
world-sheet is weighted with     
\begin{equation}\label{thoo1}
 (N\lambda_s)^b \lambda_s^c \lambda^{2g-2}_s=\lambda^{2g-2+b+c} N^{-c-2g+2}~.
\end{equation}
The 't Hooft's large $N$ limit then corresponds to taking the limit $N\rightarrow \infty$ with
$\lambda=N\lambda_s$ fixed, where $\lambda_s$ is the Type IIB string coupling. (Here we identify $\lambda_s=g_{YM}^2$, where $g_{YM}$ is the Yang-Mills coupling of the D3-brane gauge theory.) Note that addition of a cross-cap or a handle results in a diagram suppressed by an additional 
power of $N$, so that in the large $N$ limit the contributions of cross-caps and handles are subleading.
In fact, in \cite{zura} it was shown that for string vacua which are perturbatively consistent 
(that is, the tadpoles cancel) calculations of correlation functions 
in ${\cal N}<4$ gauge theories
reduce to the corresponding calculations in the parent ${\cal N}=4$
{\em oriented} theory. This holds not only for finite (in the large $N$ limit) gauge theories
but also for the gauge theories which are not conformal. (In the latter case the gauge coupling running was shown to be suppressed in the large $N$ limit.)

{}One distinguishing feature of large $N$ gauge theories obtained via
orientifolds is that the number
of possibilities which possess well defined world-sheet expansion (that is, are perturbative
from the orientifold viewpoint) is rather limited. 
In particular, in \cite{zura,zura1,class} (also see \cite{FS,blum}) ${\cal N}=1$ and ${\cal N}=2$
supersymmetric large $N$ gauge theories from orientifolds were studied in detail. Construction of these models is accompanied by certain subtleties. Thus, naively it might appear that the corresponding orientifolds of Type IIB on ${\bf R}^6/\Gamma$ should result in theories with well defined world-sheet description for any orbifold group $\Gamma\subset Spin(6)$. This is, however, not the case. For supersymmetric models it was shown in \cite{zura,zura1,class}
that only a small number of orbifold groups results in such theories. In all the other cases the corresponding orientifolds contain non-perturbative states (which can be viewed as arising from D-branes wrapping various collapsed two-cycles in the orbifold). Multiple independent checks \cite{KST,zura,zura1,class} have confirmed these conclusions. 

{}Although the present understanding of large $N$ supersymmetric gauge theories from orientifolds appears to be rather complete \cite{zura,zura1,FS,class}, no non-supersymmetric examples of such theories have been constructed. In fact, there are certain difficulties associated with such a construction. Two of the main reasons that make it non-trivial to construct non-supersymmetric large $N$ gauge theories from orientifolds are the following. First, just as in the supersymmetric case, {\em a priori} we expect non-perturbative contributions to the orientifold spectrum at least for some choices of the orbifold group. Checking whether such states are present in a given non-supersymmetric orientifold is much more non-trivial than in the
supersymmetric cases where one can use Type I-heterotic duality \cite{PW} along the lines
of \cite{ZK,KS2,KST,class}, as well as F-theory \cite{vafa} considerations \cite{KST}. Moreover,
in the cases without supersymmetry the corresponding orientifolds generically contain tachyons in (some of) the twisted closed string sectors. As we explain in the following, this by itself does not affect the consistency of the gauge theory in the large $N$ limit (as the closed string sector decouples). However, one potentially has twisted as well as massless tadpoles, and generically the tadpole cancellation conditions are overconstrained. 

{}In this paper we construct a class of non-supersymmetric large $N$ gauge theories from orientifolds in which all the tadpoles (that is, both massless and tachyonic ones) are cancelled. This is achieved by starting with ${\cal N}=1,2$ supersymmetric orientifolds of Type IIB on ${\bf C}^3/\Gamma$
with well-defined world-sheet expansion, and considering non-supersymmetric orientifolds of Type IIB on ${\bf C}^3/\Gamma^\prime$, where $\Gamma^\prime$ is obtained via a modification
of the action of $\Gamma$. This modification amounts to including {\em discrete torsion} such that a ${\bf Z}_2$ subgroup of $\Gamma^\prime$ acts differently on space-time bosonic and fermionic sectors of the orientifold. This way we obtain non-trivial anomaly free non-supersymmetric large $N$ gauge theories. In particular, we find both chiral and non-chiral gauge theories. In the former case there are certain subtleties (related to possible non-perturbative states) which we explain in detail in section IV.  

{}The remainder of this paper is organized as follows. In section II we review
some of the important points in the orientifold construction, and explain how non-trivial discrete torsion breaks supersymmetry. In section III we construct non-chiral models. In section IV we give a construction of chiral models. There we also point out certain subtleties that arise in the construction of these models. In section V we give our conclusions.

\section{Preliminaries}

{}In this section we review the setup in \cite{zura} which leads to supersymmetric 
large $N$ gauge theories from orientifolds. We then discuss how to obtain non-supersymmetric large $N$ gauge theories from orientifolds that are free of (both massless and tachyonic) tadpoles (and, consequently, of space-time anomalies).  

\subsection{Setup}

{}Consider Type IIB string theory on ${\bf C}^3/\Gamma$ where
$\Gamma\subset SU(3)(SU(2))$ so that the resulting theory has ${\cal N}=2(4)$ 
supersymmetry in four dimensions. In the following we will use the following notations: $\Gamma=\{g_a\vert a=1,\dots,|\Gamma|\}$
($g_1=1$).
Consider the $\Omega J$ orientifold of this 
theory, where $\Omega$ is the world-sheet parity reversal, and $J$ 
is a ${\bf Z}_2$ element ($J^2=1$) acting on the complex coordinates $z_i$
($i=1,2,3$) on ${\bf C}^3$ as follows: $J z_i=-Jz_i$. The resulting theory has ${\cal N}=1(2)$
supersymmetry in four dimensions. 
 
{}Note that we have an orientifold 3-plane corresponding to the $\Omega J$
element of the orientifold group. If $\Gamma$ has
a ${\bf Z}_2$ subgroup, then we also have an orientifold 7-plane.
If we have an orientifold 7-plane we must 
introduce 8 of the corresponding D7-branes to cancel the R-R charge appropriately.
(The number 8 of D7-branes is required by the corresponding tadpole cancellation
conditions.) Note, however, that the number of D3-branes is not constrained (for the corresponding untwisted tadpoles automatically vanish in the non-compact case).

{}We need to specify the action of $\Gamma$ on the Chan-Paton factors
corresponding to the D3- and D7-branes.  
These are given by Chan-Paton matrices which we collectively refer to
as $n^\mu \times n^\mu$ matrices $\gamma^\mu_a$, where the superscript $\mu$ refers to the corresponding
D3- or D7-branes. Note that ${\mbox{Tr}}(\gamma^\mu_1)=n^\mu$ where 
$n^\mu$ is the number of D-branes labelled by $\mu$. 

{}At one-loop level there are three different sources for massless tadpoles:
the Klein bottle, annulus, and M{\"o}bius strip amplitudes. The factorization property of string theory implies that the tadpole cancellation conditions read (see, {\em e.g.}, \cite{zura}
for a more detailed discussion):
\begin{equation}\label{BC}
 B_a+\sum_\mu C^\mu_a {\mbox{Tr}}(\gamma^\mu_a)=0~.
\end{equation}
Here $B_a$ and $C^\mu_a$ are (model dependent) numerical coefficients of order 1. 

{}In the world-volume of D3-branes there lives a four dimensional ${\cal N}=1(2)$ supersymmetric gauge theory (which is obtained in the low energy, that is, $\alpha^\prime\rightarrow 0$ limit). Since the number of D3-branes is unconstrained, we can consider the large
$N$ limit of this gauge theory. In \cite{zura} (generalizing the work in \cite{BKV}) it was shown that, if for a given
choice of the orbifold group $\Gamma$ the world-sheet description for the orientifold is adequate, 
then in the large $N$ limit (with $\lambda=N\lambda_s$ fixed, where $\lambda_s$ is the Type IIB string coupling) computation of any correlation function in this gauge theory is reduced to the corresponding computation in the parent ${\cal N}=4$ supersymmetric {\em oriented} gauge theory before orbifolding and orientifolding. In particular, the 
running of the gauge coupling is suppressed in the large $N$ limit. Moreover, if \begin{equation}\label{Klein}
 {\mbox {Tr}}(\gamma^\mu_a)=0~\forall a\not=1
\end{equation}
(that is, $B_a=0$ $\forall a\not=1$), then the the one-loop $\beta$-function coefficients $b_0$
for non-Abelian gauge theories living in world-volumes of the D3-branes vanish.

\subsection{Perturbative Orientifolds}

{}The arguments of \cite{zura} that imply the above properties of D3-brane gauge theories
are intrinsically perturbative. In particular, a consistent world-sheet expansion is crucial for their
validity. It is therefore important to understand the conditions for the perturbative orientifold
description to be adequate. 

{}Naively, one might expect that any choice of the orbifold group $\Gamma\subset Spin(6)$
(note that $Spin(6)$ is the $R$-symmetry group of ${\cal N}=4$ gauge theory)
should lead to an orientifold with well defined world-sheet expansion in terms of boundaries
(corresponding to D-branes), cross-caps (corresponding to orientifold planes) and handles
(corresponding to closed string loops). This is, however, not the case \cite{KST,zura1,class}. 
In fact, the number of choices of $\Gamma$ for which such a world-sheet expansion is adequate is rather constrained. In particular, in \cite{KST,zura1,class} it was argued that for ${\cal N}=1$ 
there are only seven choices of the orbifold group leading to consistent perturbative orientifolds: 
${\bf Z}_2\otimes {\bf Z}_2$ \cite{BL}, ${\bf Z}_3$ \cite{Sagnotti}, ${\bf Z}_7$, ${\bf Z}_3\otimes {\bf Z}_3$ and ${\bf Z}_6$ \cite{KS2}, ${\bf Z}_2\otimes{\bf Z}_2\otimes {\bf Z}_3$ \cite{zk}, and
$\Delta(3\cdot 3^2)$ (the latter group is non-Abelian) \cite {class}. All the other orbifold groups
(including those considered in  \cite{Zwart,AFIV}) lead to orientifolds containing sectors which are non-perturbative from the orientifold viewpoint (that is, these sectors have no world-sheet description). These sectors can be thought of as arising from D-branes wrapping various (collapsed) 2-cycles in the orbifold. The above restrictions on the orbifold group will be important in the construction of consistent non-supersymmetric large $N$ gauge theories from orientifolds. 

\subsection{Discrete Torsion and Non-Supersymmetric Models}

{}The question we are going to address next is if we can obtain {\em non-supersymmetric} large $N$ gauge theories from orientifolds via generalizing the above construction for supersymmetric theories. Such a generalization might naively seem to be straightforward. However, there are certain subtleties here. Thus, in non-supersymmetric theories we generically have tachyons (which, in particular, will be the case in the models constructed in this paper) in the twisted closed string sectors. As we will point out in a moment, the presence of tachyons by itself does not pose a problem for the consistency of the corresponding large $N$ theories. However, if tachyons are present in the physical spectrum of the corresponding orientifold model, {\em a priori} they too contribute into the tadpoles. Moreover, the cancellation conditions for the tachyonic and massless tadpoles generically are rather different. That is, the numerical coefficients $B_a$ and $C_a$ in (\ref{BC}) corresponding to the massless and tachyonic tadpoles are generically different. This typically overconstrains the tadpole cancellation conditions, which makes it rather difficult to find tadpole free non-supersymmetric orientifolds. 

{}There is a way around the above difficulties, however. Let $\Gamma\subset SU(3)$ be an orbifold group such that it contains a ${\bf Z}_2$ subgroup. Let the generator of this ${\bf Z}_2$ subgroup be $R$. Consider now the following orbifold group: $\Gamma^\prime=\{g_a^\prime| 
a=1,\dots,|\Gamma|\}$, 
where the elements $g^\prime_a$ are the same as $g_a$ except that $R$
is replaced everywhere by $R^\prime=R T$, where $T$ is the generator of a ${\bf Z}_2$ group 
corresponding to the {\em discrete torsion}\footnote{Note that ``discrete torsion'' 
$T$ only acts on the space-time fermionic sectors, and should not be confused with discrete torsion as used in orbifolds such as ${\bf Z}_2\otimes {\bf Z}_2$. Thus, the action of $T$ in the closed string sectors can be written as $(-1)^{F_L+F_R}$, where $F_L$ and $F_R$ are the space-time fermion numbers in left- and right-moving closed string sectors, respectively.}. The action of $T$ is defined as follows: it acts as identity in the bosonic sectors (that is, in the NS-NS and R-R closed string sectors, and in the NS open string sector), and it acts as $-1$ in the fermionic sectors
(that is, in the NS-R and R-NS closed string sectors, and in the R open string sector) when, say, acting on the ground states. Now consider Type IIB on ${\bf C}^3/\Gamma^\prime$. Generically, in this theory all supersymmetries are broken. 
There are certain ``exceptions'', however. Thus,
if $\Gamma\approx{\bf Z}_2$ such that $\Gamma\subset SU(2)$, then inclusion of the discrete torsion does not break supersymmetry. Similarly, if $\Gamma\approx{\bf Z}_2\otimes {\bf Z}_2$ such that $\Gamma\subset SU(3)$, the number of unbroken supersymmetries is not affected by the discrete torsion. The basic reason for this is that the ${\bf Z}_2$ twist is self-conjugate. On the other hand, in certain cases we can include the discrete torsion in ways slightly different from the one just described. In particular, let $\Gamma\approx{\bf Z}_4$ such that $\Gamma\subset SU(2)$. Let $g$ be the generator of this ${\bf Z}_4$. Next, consider the following orbifold group:
$\Gamma^\prime=\{1,gT,g^2,g^3 T\}$. In other words, the ${\bf Z}_4$ twists $g$ and $g^3$ (but not the corresponding ${\bf Z}_2$ twist $g^2$) are accompanied by the discrete torsion $T$. In this case we also have no unbroken supersymmetries. More generally, we can include the discrete torsion if the orbifold group $\Gamma$ contains a ${\bf Z}_{2^n}$ subgroup. However, the corresponding orbifold group $\Gamma^\prime$ does not always lead to non-supersymmetric theories.    

{}Next, suppose using the above construction we have found an orbifold group $\Gamma^\prime$ such that Type IIB on ${\bf C}^3/\Gamma^\prime$ is non-supersymmetric. Then it is not difficult to show that the following statement holds. If $\Gamma$ is such that the $\Omega J$ orientifold of Type IIB on ${\bf  C}^3/\Gamma$ is a perturbatively well defined 
${\cal N}=1$ or ${\cal N}=2$ theory (that is, all the massless tadpoles cancel, and there are no non-perturbative contributions to the massless spectrum), then in the $\Omega J$ orientifold of Type IIB on ${\bf  C}^3/\Gamma^\prime$ (which is non-supersymmetric) all the tachyonic and massless one-loop tadpoles (and, consequently, all the anomalies) automatically cancel. This fact is the key observation in the construction of anomaly free non-supersymmetric large $N$ gauge theories from orientifolds which we give in the subsequent sections.

\subsection{Large $N$ Limit}

{}As we already mentioned above, in all the non-supersymmetric models constructed in this paper there are tachyons in some of the twisted closed string sectors. (These are GSO projected out in orbifolds without the discrete torsion but are kept if the discrete torsion is non-trivial.) This immediately raises a question of whether the corresponding orientifolds are meaningful. 
In particular, in the presence of tachyons we expect vacuum instability.
However, there is a subtlety here which saves the day. The point is that here we are interested in large $N$ gauge theories in the 't Hooft limit $N\lambda_s={\mbox{fixed}}$, which implies that the closed string coupling constant $\lambda_s\rightarrow 0$ as we take $N$ to infinity. In particular, all the world-sheets with handles (corresponding to closed string loops) as well as cross-caps (corresponding to orientifold planes) are suppressed in this limit. That is, the closed string sector decouples from the open string sector in this limit, and after taking $\alpha^\prime\rightarrow 0$ we can ignore the closed sting states (regardless of whether they are tachyonic or not) altogether. In other words, the string construction here is simply an efficient and fast way of obtaining a field theory result, and at the end of the day we are going to throw out all the irrelevant ingredients (such as complications in the closed string sectors due to the presence of tachyons) and keep only those relevant for the field theory discussion. Note that we would not be able to do the same had we considered a {\em compact} model (with a finite number of D3-branes). In this case the closed string sector does not decouple and the theory is sick due to the presence of tachyons. 

{}Thus, in the large $N$ limit of 't Hooft (accompanied by taking $\alpha^\prime$ to zero) the non-supersymmetric gauge theory living in the world-volumes of the D3-branes is completely well defined, and following \cite{zura} we conclude that in the non-supersymmetric gauge theories arising from the orientifolds with well defined world-sheet expansions computation of any correlation function reduces to the corresponding computation in the parent ${\cal N}=4$ oriented gauge theory (before orbifolding and orientifolding). The only remaining question is which of these non-supersymmetric orientifolds have well defined world-sheet expansion. The answer to this question should be clear from our previous discussions: as long as the supersymmetric $\Omega J$ orientifold of Type IIB on ${\bf C}^3/\Gamma$ is perturbatively well defined, we expect the corresponding non-supersymmetric $\Omega J$ orientifold of Type IIB on ${\bf C}^3/\Gamma^\prime$ to also possess a well defined world-sheet expansion. This will be our guiding principle in constructing consistent non-supersymmetric large $N$ gauge theories from orientifolds in the following sections. The fact that non-Abelian gauge anomalies cancel non-trivially in the models discussed in this paper indicates self-consistency of this assumption. (However, as we discuss in section IV, such an expectation might not hold in some cases.)    

\section{Non-Chiral ${\cal N}=0$ Gauge Theories}

{}In this section we construct non-chiral large $N$ gauge theories from non-supersymmetric Type IIB orientifolds. The idea here is 
the following. Consider Type IIB on ${\bf C}\otimes ({\bf C}^2/\Gamma$)
where $\Gamma\subset SU(2)$. Suppose now
$\Gamma$ contains a ${\bf Z}_2$ subgroup. If there is no discrete torsion accompanying its generator in the fermionic sectors, then the theory will be ${\cal N}=2$ supersymmetric (after orientifolding). However, if we include non-trivial discrete torsion, the supersymmetry is going to be broken. Actually, if $\Gamma\approx {\bf Z}_2$ (such that $\Gamma\subset SU(2)$), then
(as we already mentioned in section II)  including the discrete torsion cannot break supersymmetry. On the other hand, as was shown in \cite{zura} in detail, only $\Gamma\approx{\bf Z}_M$, $M=2,3,4,6$ orbifold groups lead to perturbatively well defined $\Omega J$ orientifolds of Type IIB on ${\bf C}\otimes ({\bf C}^2/\Gamma$)
without discrete torsion (and such orientifolds have ${\cal N}=2$ supersymmetry). In all the other cases some of the massless tadpoles cannot be cancelled. We will therefore concentrate on these orbifold groups. The cases $M=2,3$ are of no interest to us: for $M=2$, as we just discussed, inclusion of the discrete torsion does not break supersymmetry; in the $M=3$ case we have no ${\bf Z}_2$ subgroup, hence we cannot have discrete torsion. The only cases left then are the ${\bf Z}_4$ and ${\bf Z}_6$ cases. 

{}We are now ready to give an explicit construction of large $N$ gauge theories from the above non-supersymmetric orientifolds.

\subsection{The ${\bf Z}_6$ orbifold}      

{}Let $g$ and $R$ be the generators of the ${\bf Z}_3$ and ${\bf Z}_2$ subgroups of the orbifold
group $\Gamma\approx{\bf Z}_6\approx{\bf Z}_3\otimes{\bf Z}_2$. The action of $g$ and $R$ on the complex coordinates $z_s$ is given by:
\begin{eqnarray}
 &&g z_1=z_1~,~~~g z_2= \omega z_2~,~~~g z_3= \omega^{-1}z_3~,~~~\omega=(2\pi i/3)~,\\
 &&R z_1=z_1~,~~~R z_2=-  z_2~,~~~R z_3= -z_3~.
\end{eqnarray} 
Now consider the orbifold group $\Gamma^\prime$ where the ${\bf Z}_2$ twist is accompanied by non-trivial discrete torsion, that is, the generator $R$ is replaced by $RT$. Supersymmetry is broken completely in this case.

{}In this model we have $n_3$ D3-branes, and 8 D7-branes. The world-volumes of the D3-branes fill the non-compact space ${\bf R}^4$ transverse to the coordinates $z_s$. 
The world-volumes of the D7-branes fill the non-compact space transverse to the $z_1$ coordinate.
The solution to the twisted tadpole cancellation conditions is given by
($N=(n_3-2)/6$):
\begin{eqnarray}
 &&\gamma_{g,3}=
 {\mbox{diag}}(\omega {\bf I}_{2N},\omega^{-1} {\bf I}_{2N}, {\bf I}_{2N+2} )~,\\
 &&\gamma_{R,3}=
 {\mbox{diag}}(i ,-i)\otimes {\bf I}_{3N+1}~,\\
 &&\gamma_{g,7}=
 {\mbox{diag}}(\omega {\bf I}_{2},\omega^{-1} {\bf I}_{2},{\bf I}_4)~,\\
 &&\gamma_{R,7}=
 {\mbox{diag}}(i ,-i)\otimes {\bf I}_{4}~.
\end{eqnarray}
The massless spectrum of this model is given in Table I. 
The gauge group is $[U(N)^2\otimes U(N+1)]_{33}
\otimes [U(1)^2\otimes U(2)]_{77}$. Note that the one-loop $\beta$-function coefficients $b_0(N)$ and $b_0(N+1)$ for the $SU(N)$
and $SU(N+1)$ subgroups of the 33 sector gauge group are independent of $N$:
\begin{eqnarray}
 &&b_0(N)=-3/2~,\\
 &&b_0(N+1)=+3~.
\end{eqnarray} 

{}Note that in some of the twisted closed string sectors we have tachyons, namely, the tachyons
arise in the $gR$ and $g^{-1}R$ twisted sectors. All the other closed string sectors are tachyon free. 

\subsection{The ${\bf Z}_4$ orbifold}      

{}Let $g$ be the generator of the orbifold
group $\Gamma\approx{\bf Z}_4$. The action of $g$ 
on the complex coordinates $z_s$ is given by:
\begin{equation}
 g z_1=z_1~,~~~g z_2= i z_2~,~~~g z_3= -i z_3~.
\end{equation} 
Now consider the orbifold group $\Gamma^\prime$ where the 
${\bf Z}_4$ twists $g$ and $g^3$ are accompanied by non-trivial discrete torsion, that is, $g$ and $g^3$ are replaced by $gT$ and $g^3 T$, respectively. (Note that this discrete torsion is such that the $g^2$ twist is torsion free.) Supersymmetry is broken completely in this case.

{}In this model we have $n_3$ D3-branes, and 8 D7-branes. The world-volumes of the D3-branes fill the non-compact space ${\bf R}^4$ transverse to the coordinates $z_s$. 
The world-volumes of the D7-branes fill the non-compact space transverse to the $z_1$ coordinate.
The solution to the twisted tadpole cancellation conditions is given by
($N=n_3/4$):
\begin{eqnarray}
 &&\gamma_{g,3}=
 {\mbox{diag}}(\omega {\bf I}_{N},\omega^{-1} {\bf I}_{N}, \omega^3 {\bf I}_{N},\omega^{-3} {\bf I}_{N} )~,\\
 &&\gamma_{g,7}=
 {\mbox{diag}}(\omega {\bf I}_{2},\omega^{-1} {\bf I}_{2}, \omega^3 {\bf I}_{2},\omega^{-3} {\bf I}_{2} )~.
\end{eqnarray}
Here $\omega=\exp (\pi i /4)$.
The massless spectrum of this model is given in Table I. 
The gauge group is $[U(N)^2]_{33}
\otimes [U(2)^2]_{77}$. Note that the one-loop $\beta$-function coefficient $b_0(N)$ for each of the $SU(N)$ subgroups 
of the 33 sector gauge group is independent of $N$. In fact, this coefficient vanishes:
\begin{equation}
 b_0(N)=0~.
\end{equation} 
Thus, at the one-loop order the theory is conformal (even for finite $N$). According to \cite{zura}, this property persists to all loop orders in 't Hooft's large $N$ limit.

{}Note that in some of the twisted closed string sectors we have tachyons, namely, the tachyons
arise in the $g$ and $g^{3}$ twisted sectors. All the other closed string sectors are tachyon free.

\subsection{Comments}

{}Here some remarks are in order. The first comment is regarding the fact that in the ${\bf Z}_6$ case the one-loop $\beta$-function coefficients of the non-Abelian subgroups of the 33 open string sector gauge group are non-zero, whereas in the ${\bf Z}_4$ case they vanish. This is in accord with the observation of \cite{zura} that $b_0=0$ if all the twisted Chan-Paton matrices $\gamma_a$ ($a\not=1$) are traceless. Moreover, generically we do not expect $b_0$ coefficients to vanish unless all ${\mbox{Tr}}(\gamma_a)$ ($a\not=1$) are traceless. However, there can be ``accidental'' cancellations in some models such that all $b_0=0$ despite some of the twisted Chan-Paton matrices not being traceless. Such ``accidental'' cancellations, in particular, occur in the ${\cal N}=2$ supersymmetric ${\bf Z}_3$ and ${\bf Z}_6$ models discussed in \cite{zura}. These cancellations were explained in \cite{zura} using the results obtained in \cite{NS}. On the other hand, such an ``accidental'' cancellation does not occur in the above non-supersymmetric ${\bf Z}_6$ model.

{}The second remark is related to the following. As discussed in \cite{Pol,KST}, the orientifold projection $\Omega$ in the above cases maps the $g_a$ twisted sector to its inverse $g^{-1}_a$ twisted sector. As discussed at length in \cite{KST}, such a projection implies that there are no non-perturbative (from the orientifold viewpoint) states arising in sectors corresponding to the orientifold group elements $\Omega g_a$ and $\Omega g_a^{-1}$ for 
$g_a^2\not=1$. This property of such orientifolds is independent of the space-time supersymmetry. This is not a trivial statement as it need not hold generically, namely, it is far from being obvious in the cases we discuss in the next section.  

\section{Chiral ${\cal N}=0$ Gauge Theories}

{}In this section we construct chiral large $N$ gauge theories from 
non-supersymmetric Type IIB orientifolds. We start with Type IIB on  ${\bf C}^3/\Gamma$, where $\Gamma$ is one of the $SU(3)$ subgroups (discussed in subsection B of section II) leading to perturbative orientifolds. For us to be able to include 
non-trivial discrete torsion, $\Gamma$ must contain a ${\bf Z}_2$ subgroup. As we already mentioned in section II, including discrete torsion in the ${\bf Z}_2\otimes {\bf Z}_2$ case
does not break supersymmetry. We are therefore led to consider the ${\bf Z}_6$ and ${\bf Z}_2\otimes {\bf Z}_2 \otimes {\bf Z}_3$ cases only.   

\subsection{The ${\bf Z}_6$ orbifold}

{}Let $g$ and $R$ be the generators of the ${\bf Z}_3$ and ${\bf Z}_2$ subgroups of the orbifold
group $\Gamma\approx{\bf Z}_6\approx{\bf Z}_3\otimes{\bf Z}_2$. The action of $g$ and $R$ on the complex coordinates $z_s$ is given by:
\begin{eqnarray}
 &&g z_s=\omega z_s~,~~~\omega=\exp(2\pi i/3)~,\\
 &&R z_1=-z_1~,~~~R z_2=-  z_2~,~~~R z_3= z_3~.
\end{eqnarray} 
Now consider the orbifold group $\Gamma^\prime$ where the ${\bf Z}_2$ twist $R$ is accompanied by non-trivial discrete torsion, that is, $R$ is replaced by $RT$. Supersymmetry is broken completely in this case. 

{}In this model we have $n_3$ D3-branes, and 8 D7-branes. The world-volumes of the D3-branes fill the non-compact space ${\bf R}^4$ transverse to the coordinates $z_s$. 
The world-volumes of the D7-branes fill the non-compact space transverse to the coordinate $z_3$.
The solution to the twisted tadpole cancellation conditions is given by
($N=(n_3+4)/6$):
\begin{eqnarray}
 &&\gamma_{g,3}=
 {\mbox{diag}}(\omega {\bf I}_{2N},\omega^{-1} {\bf I}_{2N}, {\bf I}_{2N-4} )~,\\
 &&\gamma_{R,3}=
 {\mbox{diag}}(i ,-i)\otimes {\bf I}_{3N-2}~,\\
 &&\gamma_{g,7}=
 {\mbox{diag}}(\omega {\bf I}_{4},\omega^{-1} {\bf I}_{4})~,\\
 &&\gamma_{R,7}=
 {\mbox{diag}}(i ,-i)\otimes {\bf I}_{4}~.
\end{eqnarray}
The massless spectrum of this model is given in Table II. 
The gauge group is $[U(N)^2\otimes U(N-2)]_{33}
\otimes [U(2)^2]_{77}$. Note that the non-Abelian gauge anomaly is cancelled in this model. Moreover, the one-loop $\beta$-function coefficients $b_0(N)$ and $b_0(N-2)$ for the $SU(N)$
and $SU(N-2)$ subgroups of the 33 sector gauge group are independent of $N$:
\begin{eqnarray}
 &&b_0(N)=+3~,\\
 &&b_0(N-2)=-6~.
\end{eqnarray} 
Note that in the $gR$ and $g^2 R$ twisted closed string sectors there are physical tachyons. All the other closed string sectors are tachyon free.

\subsection{The ${\bf Z}_2\otimes {\bf Z}_2\otimes {\bf Z}_3$ Orbifold}

{}Let $g$, $R_1$ and $R_2$ be the generators of the ${\bf Z}_3$ and the two ${\bf Z}_2$ subgroups of the orbifold
group $\Gamma\approx{\bf Z}_2\otimes{\bf Z}_2\otimes {\bf Z}_3$. The action of $g$ and $R_s$ ($R_3=R_1R_2$) on the complex coordinates $z_{s^\prime}$ is given by (there is no summation over the repeated indices here):
\begin{eqnarray}
 &&g z_s=\omega z_s~,~~~\omega=\exp(2\pi i/3)~,\\
 &&R_s z_{s^\prime}=-(-1)^{\delta_{ss^\prime}} z_{s^\prime}~.
\end{eqnarray}
Without loss of generality we can consider the orbifold group 
$\Gamma^\prime$
where $R_1$ has no discrete torsion, whereas $R_2$ (and, therefore, $R_3$) is accompanied by non-trivial discrete torsion. (That is, $R_2$ and $R_3$ are replaced by $R_2 T$ and $R_3 T$, respectively.)

{}In this model we have $n_3$ D3-branes, and three sets of D7-branes, which we refer too as D$7_s$-branes, with 8 D7-branes in each set. The world-volumes of the D3-branes fill the non-compact space ${\bf R}^4$ transverse to the coordinates $z_s$. 
The world-volumes of the D$7_s$-branes fill the non-compact space transverse to the
coordinate $z_s$. 
The solution to the twisted tadpole cancellation conditions is given by ($N=(n_3+4)/6$):
\begin{eqnarray}
 &&\gamma_{g,3}={\mbox{diag}}({\bf W}\otimes {\bf I}_{N}, {\bf I}_{2N-4})~,\\
 &&\gamma_{R_s,3}=i\sigma_s \otimes {\bf I}_{3N-2}~.
\end{eqnarray}
Here ${\bf W}={\mbox{diag}}(\omega,\omega,\omega^{-1},\omega^{-1})$. (The action on the D$7_s$-branes is similar.)
The massless spectrum of this model is given in Table III. The gauge group is $[U(N)\otimes Sp(N-2)]_{33}
\otimes \bigotimes_{s=1}^3 [U(2)]_{7_s7_s}$. (Here we are using the convention where the rank of $Sp(2M)$ is $M$.) Note that the non-Abelian gauge anomaly is cancelled in this model. Moreover, the one-loop $\beta$-function coefficients $b_0(N)$ and $b_0(N-2)$ for the $SU(N)$
and $Sp(N-2)$ subgroups of the 33 sector gauge group are independent of $N$:
\begin{eqnarray}
 &&b_0(N)=+1~,\\
 &&b_0(N-2)=-2~.
\end{eqnarray}
Note that in the $gR_2$, $g^2 R_2$, $gR_3$ and $g^2 R_3$ twisted closed string sectors there are physical tachyons. All the other closed string sectors are tachyon free.

\subsection{Comments}

{}As we already mentioned, not all choices of the orbifold group $\Gamma\subset SU(3)$ lead to perturbatively well defined ${\cal N}=1$ supersymmetric $\Omega J$ orientifolds of Type IIB on ${\bf C}^3/\Gamma$. Let us review the reasons responsible for such a limited number of perturbative orientifolds. Thus, consider the $\Omega J$ orientifold of Type IIB on ${\bf C}^3/\Gamma$. The orientifold
group is given by ${\cal O}=\{g_a,\Omega Jg_a\vert a=1,\dots, |\Gamma|\}$. The sectors labeled by $g_a$ correspond to the unoriented closed twisted plus untwisted sectors. The sectors labeled by $\Omega Jg_a$ with $(Jg_a)^2=1$ correspond to open strings stretched between D-branes. (In particular, if the set of points fixed under $Jg_a$ has dimension 0
then these are D3-branes. If this set has real dimension 4, then these are D7-branes.) However,
the sectors labeled by $\Omega Jg_a$ with $(Jg_a)^2\not=1$ do not have an interpretation in
terms of open strings starting and ending on perturbative D-branes ({\em i.e.}, they do not have an interpretation in terms of open strings with purely Dirichlet or Neumann boundary conditions
in all directions) \cite{KST}. Instead, if viewed as open strings they would have mixed (that is, neither Dirichlet nor Neumann) boundary conditions. These states do not have world-sheet
description. They can be viewed as arising from D-branes wrapping (collapsed) two-cycles in the
orbifold \cite{KST}. These states are clearly non-perturbative from the orientifold viewpoint.

{}This difficulty is a generic feature in most of the orientifolds of Type IIB compactified on toroidal
orbifolds, as well as the corresponding non-compact cases such as the $\Omega J$ orientifolds of Type IIB on ${\bf C}^3/\Gamma$. However, there is a (rather limited) class of cases where the would-be non-perturbative
states are massive (and decouple in the low energy effective field theory) if we consider
compactifications on blown up orbifolds \cite{KST}. In fact, these blow-ups are forced by
the orientifold consistency.
The point is that the orientifold projection $\Omega$ must be chosen to be the same 
as in the case of Type IIB on a smooth Calabi-Yau three-fold. 
The reasons why this choice of the orientifold
projection is forced have been recently discussed at length in \cite{KST}. In particular, we do not
have an option of choosing the orientifold projection analogous to that in the six dimensional models of \cite{GJ}. (Instead, the orientifold projection must be analogous to 
that in the ${\bf Z}_2$ models of \cite{PS}.) On the other hand, the above $\Omega$ orientifold projection is
not a symmetry of Type IIB on ${\bf C}^3/\Gamma$ at the orbifold conformal field theory point \cite{KST}.
The reason for this is that $\Omega$ correctly reverses the world-sheet orientation of world-sheet bosonic and fermionic oscillators and left- and right-moving momenta, but fails
to do the same with the {\em twisted} ground states. (Such a reversal would involve mapping  
the $g_a$ twisted ground states to the $g^{-1}_a$ twisted ground states.
In \cite{KST} such an orientation reversal was shown to be inconsistent.) This difficulty is circumvented by noting that the orientifold projection $\Omega$ is consistent for
{\em smooth} Calabi-Yau three-folds, and, in particular, for a blown up version of the ${\bf C}^3/\Gamma$
orbifold. Thus, once the appropriate blow-ups are performed, the orientifold procedure is well
defined.

{}In some cases the blow-ups result in decoupling of the would-be massless non-perturbative
states, which is due to the presence of an appropriate superpotential (that couples the blow-up
modes to the non-perturbative states). This feature, however, is not generic and is only present
in a handful of cases. This was shown to be the case in $\Omega$ orientifolds of Type IIB on
$T^6/\Gamma$ with $\Gamma\approx{\bf Z}_3$, ${\bf Z}_7$, ${\bf Z}_3\otimes 
{\bf Z}_3$, $\Delta(3\cdot 3^2)$ in \cite{ZK,KS2,class}. These orientifolds correspond to Type I compactifications on
the corresponding orbifolds, which in turn have perturbative heterotic duals (the corresponding heterotic compactifications are perturbative as there are no D5-branes (which would map to heterotic NS5-branes) in these models). The non-perturbative (from the orientifold viewpoint) states were
shown to correspond to twisted sector states on the heterotic side. The perturbative superpotentials for these states can be readily computed, and are precisely such that after the
appropriate blow-ups (those needed for the orientifold consistency) the twisted sector states decouple.

{}These arguments were generalized in \cite{KST} to the ${\bf Z}_6$
model of \cite{KS2} and the ${\bf Z}_2\otimes {\bf Z}_2\otimes {\bf Z}_3$ model of \cite{zk}.
In all the other cases (except for the ${\bf Z}_2\otimes {\bf Z}_2$ model of \cite{BL} which is
obviously perturbative from the above viewpoint) it was argued in \cite{KST} that non-perturbative states do not decouple. Various checks of these statements were performed in
\cite{KST} using the web of dualities between Type IIB orientifolds, F-theory, and Type I and heterotic compactifications on orbifolds. These statements, however, only depend on local properties
of D-branes and orientifold planes near orbifold singularities and should persist in non-compact
cases such as the $\Omega J$ orientifolds of Type IIB on ${\bf C}^3/\Gamma$. 
In \cite{zura1} it was shown that only for the above seven choices of the orbifold group do the
perturbative (from the orientifold viewpoint) tadpole cancellation conditions have appropriate solutions for the
$\Omega J$ orientifolds of Type IIB on ${\bf C}^3/\Gamma$.

{}The reason we have reviewed the above facts is the following. In the non-supersymmetric cases we discussed in this section {\em a priori} we also might expect non-perturbative states arising in various sectors of the orientifold. Unlike in the supersymmetric cases, however, it is unclear whether such states would decouple once the orbifold singularities are resolved. In fact, it is not even clear if the corresponding ``blow-ups'' are marginal deformation (since supersymmetry is broken). In particular, we do not have the dual heterotic picture in these cases which we would use to check the decoupling of non-perturbative (from the orientifold viewpoint) states: such heterotic duals would be intrinsically non-perturbative (as they would contain NS 5-branes), and it is not clear how to proceed in these cases. Thus, we do not really have an independent check in the non-supersymmetric cases (in contrast to the ${\cal N}=1$ supersymmetric cases) for the perturbative consistency of the non-compact orientifolds we constructed in this section. However, non-Abelian anomaly cancellation in these models is rather non-trivial, so it is reasonable to believe that these models are indeed perturbatively consistent. Yet, the above discussion points to a possible caveat in the above construction.

\section{Conclusions}

{}In this paper we have constructed non-supersymmetric large $N$ gauge theories from orientifolds. The construction is similar to that of the supersymmetric models but involves non-trivial discrete torsion which is the source of supersymmetry breaking. The non-chiral models we have constructed in this paper are consistent as they are obtained by including non-trivial discrete torsion in ${\cal N}=2$ theories (in which we do not expect non-perturbative states due to a peculiar orientifold projection). However, the situation with the chiral models is less clear: they are obtained by including non-trivial discrete torsion in ${\cal N}=1$ theories where {\em a priori} we do expect non-perturbative states. Unlike in the supersymmetric cases, in the non-supersymmetric case decoupling of such states is far from being obvious. It would be interesting to understand this issue in more detail. However, assuming that such states do decouple, the chiral models (along with the non-chiral ones) we have constructed in this paper provide examples of non-supersymmetric large $N$ gauge theories from orientifolds with well defined world-sheet expansion which is in one-to-one correspondence with 't Hooft's large $N$ expansion (and results in rather non-trivial statements about the corresponding gauge theories in the large $N$ limit).

{}In conclusion we would like to stress that if we attempted to construct {\em compact} models using the above techniques, we would get tachyonic models in which the closed string sector (unlike in the large $N$ limit of the non-compact cases) does not decouple from the gauge theory, and these models would be sick due to tachyonic instabilities. It remains an open question whether it is possible to construct chiral tachyon and tadpole (and, therefore, anomaly)
free compact Type I models. It would be interesting to understand this issue in more detail.

\acknowledgments

{}This work was supported in part by the grant NSF PHY-96-02074, 
and the DOE 1994 OJI award. I would also like to thank Albert and Ribena Yu for 
financial support.

\begin{table}[t]
\begin{tabular}{|c|c|c|l|}
 Model &Gauge Group  & Charged Bosons &Charged Fermions
  \\
 \hline
&&&\\
${\bf Z}_6$ & $[U(N)^2\otimes U(N+1)]_{33}\otimes$  &  
 $3\times {1\over 2} [({\bf 1},{\bf 1},{\bf 1};{\bf 1})_b]_{33}$ &
 $[({\bf N},{\overline {\bf N}},{\bf 1};{\bf 1})_f]_{33}$
 \\
            &  $[U(1)^2\otimes U(2)]_{77}$ 
 &
 ${1\over 2} [({\bf Adj},{{\bf 1}},{\bf 1};{\bf 1})_b]_{33}$ & 
 $[({\overline {\bf N}},{\bf N},{\bf 1};{\bf 1})_f]_{33}$\\
           &  & ${1\over 2} [({{\bf 1}},{\bf Adj},{\bf 1};{\bf 1})_b]_{33}$ &
 $[({ {\bf 1}},{\bf 1},{\bf A};{\bf 1})_f]_{33}$ \\
           &  & ${1\over 2} [({{\bf 1}},{\bf 1},{\bf Adj};{\bf 1})_b]_{33}$ &
 $[({ {\bf 1}},{\bf 1},{\overline {\bf A}};{\bf 1})_f]_{33}$ \\
            &   &
 $[({\bf A},{{\bf 1}},{\bf 1};{\bf 1})_b]_{33}$ & 
     $[({ {\bf N}},{\bf N},{ {\bf 1}};{\bf 1})_f]_{33}$\\
           &  &  $[({{\bf 1}},{\bf A},{\bf 1};{\bf 1})_b]_{33}$ &
 $[({ {\bf N}},{\bf 1},{ \overline {\bf N+1}};{\bf 1})_f]_{33}$\\
            &   &
 $[({\bf N},{{\bf 1}},{\bf N+1};{\bf 1})_b]_{33}$ &
  $[({ {\bf 1}},{\bf N},{{\bf N+1}};{\bf 1})_f]_{33}$ \\
           &  &  $[({{\bf 1}},{\bf N},{\overline {\bf N+1}};{\bf 1})_b]_{33}$ & \\
  &   &  
 ${1\over 2} [({\bf 1},{\bf 1},{\bf 1};{\bf 3})_b]_{77}$ &
 \\
 &   &  
 $3\times {1\over 2} [({\bf 1},{\bf 1},{\bf 1};{\bf 1})_b]_{77}$ &
 $5\times [({\bf 1},{\bf 1},{\bf 1};{\bf 1})_f]_{77}$
 \\
 &   &  
 $2\times  [({\bf 1},{\bf 1},{\bf 1};{\bf 2})_b]_{77}$ & $2\times [({\bf 1},{\bf 1},{\bf 1};{\bf 2})_f]_{77}$
 \\
 &   &  
 ${1\over 2} [({\bf N},{\bf 1},{\bf 1};{\bf 1})_b]_{37}$ & 
 $[({\bf N},{\bf 1},{\bf 1};{\bf 1})_f]_{37}$
 \\
&   &  
 ${1\over 2} [({\bf N},{\bf 1},{\bf 1};{\bf 2})_b]_{37}$ &
 $[({\bf 1},{\bf N},{\bf 1};{\bf 1})_f]_{37}$
 \\
 &   &  
 ${1\over 2} [({\bf 1},{\bf N},{\bf 1};{\bf 1})_b]_{37}$ &
 $[({\bf 1},{\bf 1},{\bf N+1};{\bf 2})_f]_{37}$
 \\
&   &  
 ${1\over 2} [({\bf 1},{\bf N},{\bf 1};{\bf 2})_b]_{37}$ &
 \\
&   &  
 $2\times {1\over 2} [({\bf 1},{\bf 1},{\bf N+1};{\bf 1})_b]_{37}$ &
 \\ 
\hline
&&&\\
 ${\bf Z}_4$ & $[U(N)^2]_{33}\otimes [U(2)^2]_{77}$  &  
 $[({\bf 1},{\bf 1};{\bf 1},{\bf 1})_b]_{33}$ &
 
 \\ 
&   &  
 ${1\over 2}[({\bf Adj},{\bf 1};{\bf 1},{\bf 1})_b]_{33}$ &
 $[({\bf N},{\overline {\bf N}};{\bf 1},{\bf 1})_f]_{33}$
 \\
 &   &  
 ${1\over 2}[({\bf 1},{\bf Adj};{\bf 1},{\bf 1})_b]_{33}$ &
 $[({\overline {\bf N}},{\bf N};{\bf 1},{\bf 1})_f]_{33}$
 \\
 &   &  
 $[({\bf A},{\bf 1};{\bf 1},{\bf 1})_b]_{33}$ &
 $[({ {\bf A}},{\bf 1};{\bf 1},{\bf 1})_f]_{33}$
 \\
 &   &  
 $[({\bf 1},{\bf A};{\bf 1},{\bf 1})_b]_{33}$ &
 $[({ {\bf 1}},{\bf A};{\bf 1},{\bf 1})_f]_{33}$
 \\
 &   &  
 $[({\bf N},{\bf N};{\bf 1},{\bf 1})_b]_{33}$ &
 $[({ {\bf N}},{\bf N};{\bf 1},{\bf 1})_f]_{33}$
 \\
 &   &  
 $3\times [({\bf 1},{\bf 1};{\bf 1},{\bf 1})_b]_{77}$ &
 $2\times [({ {\bf 1}},{\bf 1};{\bf 1},{\bf 1})_f]_{77}$
 \\
 &   &  
 ${1\over 2}[({\bf 1},{\bf 1};{\bf 3},{\bf 1})_b]_{77}$ &
 
 \\
&   &  
 ${1\over 2}[({\bf 1},{\bf 1};{\bf 1},{\bf 3})_b]_{77}$ &
 
 \\
&   &  
 $[({\bf 1},{\bf 1};{\bf 2},{\bf 2})_b]_{77}$ &
 $3\times [({ {\bf 1}},{\bf 1};{\bf 2},{\bf 2})_f]_{77}$
\\
&   &  
 $[({\bf N},{\bf 1};{\bf 2},{\bf 1})_b]_{37}$ &
 $[({ {\bf N}},{\bf 1};{\bf 2},{\bf 1})_f]_{37}$
 \\
&   &  
 $[({\bf 1},{\bf N};{\bf 1},{\bf 2})_b]_{37}$ &
 $[({ {\bf 1}},{\bf N};{\bf 1},{\bf 2})_f]_{37}$
 \\
\hline
\end{tabular}
\caption{The massless open string spectra of the ${\cal N}=0$ orientifolds of Type IIB on ${\bf C}\otimes ({\bf C}^2/{\bf Z}_6)$ and Type IIB on ${\bf C}\otimes ({\bf C}^2/{\bf Z}_4)$. The subscript ``$b$'' indicates that the corresponding field consists of the bosonic content of a hypermultiplet. (Thus, ${1\over 2}$ of this content corresponds to a complex scalar.)
The subscript ``$f$'' indicates that the corresponding field consists of the fermionic content
of a hypermultiplet (that is, of one left-handed and one right-handed chiral fermion in the corresponding representation of the gauge group). The notation ${\bf A}$ stands for the two-index antisymmetric
representation of the corresponding unitary group, whereas ${\bf Adj}$ stands for the adjoint representation. For the sake of simplicity we have suppressed the $U(1)$ charges (which are not
difficult to restore).}
\label{Z6Z4} 
\end{table}

\begin{table}[t]
\begin{tabular}{|c|c|l|}
 Model &Gauge Group   & Charged Complex Bosons 
  \\
 \hline
${\bf Z}_6$ & $[U(N)\otimes U(N)\otimes U(N-2)]_{33}\otimes$  &  
 $2\times [({\bf A},{\bf 1},{\bf 1};{\bf 1},{\bf 1})(+2,0,0;0,0)_c]_{33}$
 \\
            &  $[U(2)\otimes U(2)]_{77}$ 
  &
 $2\times [({\bf 1},{\overline {\bf A}},{\bf 1};{\bf 1},{\bf 1})(0,-2,0;0,0)_c]_{33}$  \\
            & 
 &
 $2\times [({\overline {\bf N}},{\bf 1},{\overline {\bf N-2}};
     {\bf 1},{\bf 1})(-1,0,-1;0,0)_c]_{33}$  \\
         & 
 &
 $2\times [({\bf 1},{\bf N},{\bf N-2};
     {\bf 1},{\bf 1})(0,+1,+1;0,0)_c]_{33}$  \\
 & &
  $[({\overline {\bf N}},{\bf 1},{\bf N-2};
     {\bf 1},{\bf 1})(-1,0,+1;0,0)_c]_{33}$  \\
          & 
 &
 $[({\bf 1},{\bf N},{\overline {\bf N-2}};
     {\bf 1},{\bf 1})(0,+1,-1;0,0)_c]_{33}$  \\
 & &$ [({\bf N},{\overline {\bf N}},{\bf 1};
     {\bf 1},{\bf 1})(+1,-1,0;0,0)_c]_{33}$  \\
  &   &  

 $2\times [({\bf 1},{\bf 1},{\bf 1};{\bf 1},{\bf 1})(0,0,0;+2,0)_c]_{77}$
 \\
            &   
 &
 $2\times [({\bf 1},{\bf 1},{\bf 1};{\bf 1},{{\bf 1}})(0,0,0;0,-2)_c]_{77}$  \\
 & &$  [(
     {\bf 1},{\bf 1},{\bf 1};{\bf 2},{{\bf 2}})(0,0,0;+1,-1)_c]_{77}$  \\
&         & $[({\bf N},{\bf 1},{\bf 1};{\bf 1},{\bf 2})
(+1,0,0;0,+1)_c]_{37}$\\
&         & $[({\bf 1},{\bf N},{\bf 1};{\bf 2},{\bf 1})
(0,+1,0;+1,0)_c]_{37}$\\
&         & $[({\overline{\bf N}},{\bf 1},{\bf 1};{\bf 1},{{\bf 2}})
(-1,0,0;0,-1)_c]_{37}$\\
&         & $[({\bf 1},{\overline {\bf N}},{\bf 1};{{\bf 2}},{\bf 1})
(0,-1,0;-1,0)_c]_{37}$\\
\hline
 & & Charged Chiral Fermions 
  \\
 \hline
 &   &
 $[({\bf A},{\bf 1},{\bf 1};{\bf 1},{\bf 1})(+2,0,0;0,0)_L]_{33}$
 \\
            &   
 &
 $[({\bf 1},{\overline {\bf A}},{\bf 1};{\bf 1},{\bf 1})(0,-2,0;0,0)_L]_{33}$  \\
            & 
 &
 $[({\overline {\bf N}},{\bf 1},{\overline {\bf N-2}};
     {\bf 1},{\bf 1})(-1,0,-1;0,0)_L]_{33}$  \\
          & 
 &
 $[({\bf 1},{\bf N},{\bf N-2};
     {\bf 1},{\bf 1})(0,+1,+1;0,0)_L]_{33}$  \\
 & &
  $2\times [({\overline {\bf N}},{\bf 1},{\bf N-2};
     {\bf 1},{\bf 1})(-1,0,+1;0,0)_L]_{33}$  \\
          & 
 &  
 $2\times [({\bf 1},{\bf N},{\overline {\bf N-2}};
     {\bf 1},{\bf 1})(0,+1,-1;0,0)_L]_{33}$  \\
 & &$2\times  [({\bf N},{\overline {\bf N}},{\bf 1};
     {\bf 1},{\bf 1})(+1,-1,0;0,0)_L]_{33}$  \\
  & &$ [({\bf N},{{\bf N}},{\bf 1};
     {\bf 1},{\bf 1})(+1,+1,0;0,0)_L]_{33}$  \\
 & &$ [({\overline {\bf N}},{\overline {\bf N}},{\bf 1};
     {\bf 1},{\bf 1})(-1,-1,0;0,0)_L]_{33}$  \\ 
   & &$ [({\bf 1},{{\bf 1}},{\bf A};
     {\bf 1},{\bf 1})(0,0,+2;0,0)_L]_{33}$  \\
 & &$ [({{\bf 1}},{{\bf 1}},{\overline {\bf A}};
     {\bf 1},{\bf 1})(0,0,-2;0,0)_L]_{33}$  \\
&   &  
 $[({\bf 1},{\bf 1},{\bf 1};{\bf 1},{\bf 1})(0,0,0;+2,0)_L]_{77}$
 \\
           &   
 &  
 $[({\bf 1},{\bf 1},{\bf 1};{\bf 1},{{\bf 1}})(0,0,0;0,-2)_L]_{77}$  \\
 & &$2\times   [(
     {\bf 1},{\bf 1},{\bf 1};{\bf 2},{{\bf 2}})(0,0,0;+1,-1)_L]_{77}$  \\
  & &$ [({\bf 1},{\bf 1},{\bf 1};{\bf 2},{{\bf 2}}
     )(0,0,0;+1,+1)_L]_{77}$  \\
 & &$ [({\bf 1},{\bf 1},{\bf 1};
 {{\bf 2}},{{\bf 2}}
     )(0,0,0;-1,-1)_L]_{77}$  \\     
&         & $[({\overline {\bf N}},{\bf 1},{\bf 1};{ {\bf 2}},{\bf 1})
(-1,0,0;-1,0)_L]_{37}$\\
& & $[({\bf 1},{\bf 1},{\overline {\bf N-2}};{\bf 1},{{\bf 2}})
(0,0,-1;0,-1)_L]_{37}$\\
 &         & $[({\bf 1},{\bf N},{\bf 1};{\bf 1},{\bf 2})(0,+1,0;0,+1)_L]_{37}$  \\ 
 & & $[({\bf 1},{\bf 1},{{\bf N-2}};{{\bf 2}},{\bf 1})
(0,0,+1;+1,0)_L]_{37}$\\
\end{tabular}
\caption{The massless open string spectrum of the ${\cal N}=0$ orientifold of Type IIB on ${\bf C}^3/
{\bf Z}_6$. The subscript ``$c$'' indicates that the corresponding field is a complex boson.
The subscript ``$L$'' indicates that the corresponding field is a left-handed chiral
fermion. The notation ${\bf A}$ stands for the two-index antisymmetric
representation of the corresponding unitary group. The $U(1)$ charges are
given in parentheses.}
\label{Z6bfN} 
\end{table}

\begin{table}[t]
\begin{tabular}{|c|c|l|}
 Model &Gauge Group   & Charged Complex Bosons 
  \\
 \hline
${\bf Z}_2\otimes{\bf Z}_2\otimes {\bf Z}_3$ & $[U(N)\otimes Sp(N-2)]_{33}\otimes$  &  
 
 $3\times [({\bf A},{\bf 1})(+2)_c]_{33}$
 \\
           &  $\bigotimes_{s=1}^3 [U(2)]_{7_s 7_s}$ 
 &  
 $3\times [({\overline {\bf N}},{{\bf N-2}})(-1)_c]_{33}$  \\
   &   &  
 
 $3\times [({\bf 1}_s)(+2_s)_c]_{7_s7_s}$
 \\
            &   
 &  
 $[({\bf N},{\bf 1};{\bf 2}_1)(+1;+1_1)_c]_{37_1}$  \\
            &   
 &  
 $[({\bf 1},{{\bf N-2}};{{\bf 2}}_1)(0;-1_1)_c]_{37_1}$  \\
 &&  
 $[({\bf 2}_2;{\bf 2}_{3}) (+1_2;+1_{3})_c]_{7_2
 7_{3}}$  \\
            &   
 &  
 $[({\bf N},{\bf 1};{{\bf 2}}_k)(+1;-1_k)_c]_{37_k}$  \\
          &   
 &  
 $[({\overline {\bf N}},{\bf 1};{\bf 2}_k)(-1;+1_k)_c]_{37_k}$  \\
          &   
 & 
 $[({\bf 2}_1;{{\bf 2}}_k)(+1_1;-1_k)_c]_{7_1 7_k}$  \\
\hline
 &&  Charged Chiral Fermions 
  \\
 \hline
 &  &  
 $ [({\bf Adj},{\bf 1})(0)_L]_{33}$
 \\ 
            &   
 &
 $ [({\bf 1},{\bf a})(0)_L]_{33}$
 \\
 &&
 $2\times [({\bf 1},{\bf 1})(0)_L]_{33}$
 \\
  &   &  
 $ [({\bf S},{\bf 1})(+2)_L]_{33}$
 \\
&   &  
 $2\times [({\bf A},{\bf 1})(+2)_L]_{33}$
 \\
            &   
 & 
 $3\times [({\overline {\bf N}},{{\bf N-2}})(-1)_L]_{33}$  \\
  &   &  
 $ [({\bf 3}_s)(+2_s)_L]_{7_s7_s}$
 \\
  &   &  
 $2\times [({\bf 1}_s)(+2_s)_L]_{7_s7_s}$
 \\
 &   &  
 $ [({\bf 3}_s)(0_s)_L]_{7_s 7_s}$
 \\ 
 & &
 $2\times [({\bf 1}_s)(0)_L]_{7_s 7_s}$
 \\
            &   
 &  
 $[({\bf N},{\bf 1};{\bf 2}_1)(+1;+1_1)_L]_{37_1}$  \\
            &   
 &  
 $[({\bf 1},{{\bf N-2}};{{\bf 2}}_1)(0;-1_1)_L]_{3 7_1}$  \\
 &&  
 $[({\bf 2}_2,{\bf 1}_2;{\bf 2}_{3},{{\bf 1}}_{3}) (+1_2;+1_{3})_L]_{7_2
 7_{3}}$  \\
            &   
 &  
 $[({\overline {\bf N}},{\bf 1};{{\bf 2}}_k)(-1;-1_k)_L]_{37_k}$  \\
            &   
 &  
 $[({\bf 1},{{\bf N-2}};{{\bf 2}}_k)(0;+1_k)_L]_{37_k}$  \\
            &   
 &  
 $[({\bf 2}_1,{\bf 2}_k)(-1_1;-1_k)_L]_{7_1 7_k}$  \\
 \hline
\end{tabular}
\caption{The massless open string spectrum of the ${\cal N}=0$ orientifold of Type IIB on ${\bf C}^3/{\bf Z}_2\otimes{\bf Z}_2\otimes {\bf Z}_3$.
The subscript ``$c$'' indicates that the corresponding field is a complex boson.
The subscript ``$L$'' indicates that the corresponding field is a left-handed chiral
fermion. The notation ${\bf A}$ stands for the two-index antisymmetric representation of $SU(N)$, whereas ${\bf S}$ stands for the two-index symmetric representation of $SU(N)$. 
The notation ${\bf Adj}$ stands for the $N^2-1$ dimensional adjoint representation of $SU(N)$,
whereas ${\bf a}$ stands for the $N(N-1)/2-1$ dimensional {\em traceless} antisymmetric representation of $Sp(N)$.
Also, $s=1,2,3$, and $k=2,3$.
The $U(1)$ charges are
given in parentheses.}
\label{Z223fN} 
\end{table}


\end{document}